\documentclass[12pt]{article}
\usepackage{graphicx,epsfig}
\def\lsim{\mathrel{\raise.3ex\hbox{$<$\kern-.75em\lower1ex\hbox{$\sim$}}}}
\def\gsim{\mathrel{\raise.3ex\hbox{$>$\kern-.75em\lower1ex\hbox{$\sim$}}}}
\newcommand{ \slashchar }[1]{\setbox0=\hbox{$#1$}   % set a box for #1
   \dimen0=\wd0                                     % and get its size
   \setbox1=\hbox{/} \dimen1=\wd1                   % get size of /
   \ifdim\dimen0>\dimen1                            % #1 is bigger
      \rlap{\hbox to \dimen0{\hfil/\hfil}}          % so center / in box
      #1                                            % and print #1
   \else                                            % / is bigger
      \rlap{\hbox to \dimen1{\hfil$#1$\hfil}}       % so center #1
      /                                             % and print /
   \fi}                                             %
\def\be{\begin{equation}}
\def\ee{\end{equation}}
\def\bea{\begin{eqnarray}}
\def\eea{\end{eqnarray}}
\def\bec{\begin{center}}
\def\eec{\end{center}}
\def\atversim#1#2{\lower0.7ex\vbox{\baselineskip\zatskip\lineskip\zatskip
  \lineskiplimit 0pt\ialign{$\matth#1\hfil##\hfil$\crcr#2\crcr\sim\crcr}}}

\renewcommand{\thefootnote}{\fnsymbol{footnote}}

\newcounter{appendixc}
\newcounter{subappendixc}[appendixc]
\newcounter{subsubappendixc}[subappendixc]

\renewcommand{\appendix}[1] {\vspace*{0.6cm}
        \refstepcounter{appendixc}
        \setcounter{figure}{0}
        \setcounter{table}{0}
        \setcounter{equation}{0}
        \renewcommand{\thefigure}{\Alph{appendixc}.\arabic{figure}}
        \renewcommand{\thetable}{\Alph{appendixc}.\arabic{table}}
        \renewcommand{\theappendixc}{\Alph{appendixc}}
        \renewcommand{\theequation}{\Alph{appendixc}.\arabic{equation}}
        \noindent{\bf Appendix \theappendixc #1}\par\vspace*{0.4cm}}

\begin{document}
\begin{titlepage}
\rightline{\vbox{\halign{&#\hfil\cr \cr
%&hep-ph/0510228
\cr }}} \vskip .5in
\begin{center}

{\Large\bf One Explanation for the Exotic State Y(4260)}

\vskip .5in \normalsize {\bf Cong-Feng Qiao
\footnote{Email: qiaocf@gucas.ac.cn}}\\
\vskip .5cm

CCAST(World Lab.), P.O. Box 8730, Beijing 100080, China\\

\vskip .3cm  Dept. of Physics, Graduate School, the Chinese
Academy of Sciences\\
YuQuan Road 19A, Beijing 100049, China \vskip 2.3cm
\end{center}
\begin{abstract}
\normalsize

In this Letter we interpret the Y(4260), a state recently
discovered by the BaBar Collaboration that has a mass within the
range of conventional charmonium states, as having a
molecular-state structure. In our scheme this molecular-like state
is not constructed out of two-quark mesons, but rather out of
baryons, i.e., the Y(4260) is a baryonium state. With this
interpretation, the  unusual measured properties of the Y(4260)
are easily understood and some further peculiar decay
characteristics are predicted.

\end{abstract}
\vspace{1cm} PACS number(s): 12.39Mk, 12.39.Pn, 13.30.Eg

\renewcommand{\thefootnote}{\arabic{footnote}}
\end{titlepage}

%%%%%%%%%%%%%%%%%%%%%%%%%%%%%%%%%%%%%%%%%%%%%%%%%%%%%%%%%%%%%%%%%%
Physics research at heavy-flavor energy scales, e.g., at charm and
bottom masses, has recently made several discoveries. Among these
are the observations of several new resonances or resonant
structures. Very recently, a resonant structure named the Y(4260)
was observed by the BaBar Collaboration in the initial-state
radiation process $e^+ e^- \rightarrow \gamma~ \pi^+ \pi^- J/\psi$
\cite{Babar}. This state must have the same quantum numbers as the
photon, $J^{PC} = 1^{--}$, and is a broad resonance with a fitted
width of
\begin{eqnarray}
\Gamma= 90\;\mathrm{MeV}\; .
\end{eqnarray}

Since the mass of the Y(4260) is within the range of conventional
charmonium states, a natural explanation is that it contains
charm-anticharm constituents. However, although the observed
enhancement is at $4.26$ GeV, quite a ways about open-charm
threshold, there is no experimental signal for $D \bar{D}$ in its
observed decays. Presently, except for Ref.\cite{felipe} where the
Y(4260) is still interpreted as a normal member in the charmonium
spectra, a common belief is that it is rather an exotic (or
cryptoexotic) state. The authors of Refs. \cite{zhu,kou,close}
propose that the new state is a charmonium hybrid that is
configured out of a pair of charm-anticharm quarks and a gluon.
Ref. \cite{maiani} thinks the Y(4260) may be the first orbital
excitation of a diquark-antidiquark state $[cs][\bar{c}\bar{s}]$.
In Ref. \cite{li}, Liu {\it et al.} explain the resonance as a
molecular state composed of a $\rho$ and $\chi_c$, while the
authors of Ref. \cite{yuan} take it as a $\omega$ and $\chi_c$
molecular state.

In our understanding, the Y(4260) might be a baryonium state
containing hidden charm and made out of
$\Lambda_c$-$\bar{\Lambda}_c$ (we use
$\Lambda_c$-$\bar{\Lambda}_c$ to represent
$\Lambda_c^+$-$\Lambda_c^-$ hereafter in this paper). This model
not only gives a natural explanation for the state's observed and
unobserved decays, but can also make predictions of some further
peculiar properties, which could be measured by future
experiments, as explained in the following.

At an $e^+\; e^-$ collider, any directly produced state is naively
thought to be flavor symmetric in its inner structure, a view held
by most of the earlier speculations on the Y(4260), e.g., the
$[cs][\bar{c}\bar{s}]$ interpretation. Our model follows this idea
as well. In the $\Lambda_c$-$\bar{\Lambda}_c$ assignment, the
$\pi^+\pi^-J/\psi$ decay process can proceed easily, as shown
schematically in Figure 1(a). However, the $D$ $\bar{D}$ exclusive
decay is almost impossible. One way to produce a $D$ $\bar{D}$
pair is to let one pair of light quarks annihilate into a
photon(Figure 1(b)), which is an electromagnetic process and,
hence, is highly suppressed relative to the observed decay mode. A
rough calculation tells us that the $D$ $\bar{D}$ $\gamma$ decay
process should be two orders of magnitude smaller than the
observed  $\pi^+\pi^-J/\psi$ decay. Another possibility is to let
a pair of light quarks annihilate into a gluon, followed by a
complicated color rearrangement to make the $D$ $\bar{D}$
configuration. Normally, this kind of decay scheme is suppressed
by the color factor $N_c$. In the baryonium model, the most
important open-charm decay process of the Y(4260) should be into
$D^* \bar{D}^* \pi$, but not into $D \bar{D} \pi$ because of the
requirements of parity, angular momentum and isospin conservation.
Since for $\Lambda_c$-$\bar{\Lambda}_c$ baryonium there are no
constituent strange quarks, the decay to final states with
strangeness, such as $D_{s}\; \bar{D_{s}}$ and $K^+\; K^-\;
J/\psi$, should also be suppressed relative to the observed mode
$\pi^+\pi^-J/\psi$\footnote{This prediction is roughly confirmed
by the recent measurement from the CLEO Collaboration \cite{cleo}.
They find evidence for a $K^+ K- J/\psi$ signal in Y(4260) decays
with a rate much smaller than for $\pi^+ \pi^- J/\psi$.}.
\begin{figure}[htdp]
\begin{center}
\psfig{figure=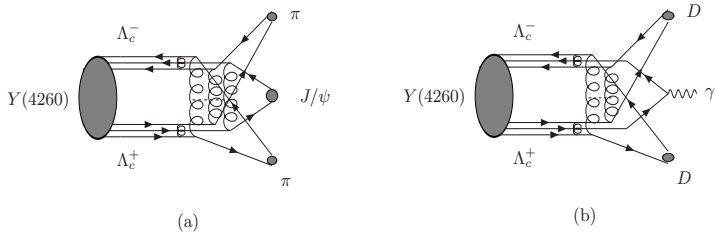,width=15cm,height=23cm
%angle=-90
} \vspace{-18.5cm} \caption{(a)The schematic Feynman diagram of
$Y(4260)$ decaying into $J/\psi$ $\pi$$\pi$; (b)The schematic
Feynman diagram of $Y(4260)$ radiative decay to two $D$ mesons and
a photon process.} \label{fig1}
\end{center}
\end{figure}
\par
In fact, the baryonium conjecture is not really a big surprise. In
the framework of QCD, this kind of states might exist, similar to
Helium. After the advent of QCD, indeed, there have been lots of
theoretical speculations and experimental efforts on it.
Nevertheless, no definite observation is confirmed to be a
baryonium up to now. With experiment in progress, to find the
signature of baryonium state by the today's highly efficient
machines is not impossible. For instance, the recent measurements
by BES \cite{BES} and BELLE \cite{Belle} stimulate a new round of
interest in baryonium physics. The BES observation on the $p
\bar{p}$ threshold enhancement in $J/\psi$ decays has been
interpreted as possibly a proton-antiproton baryonium \cite{zhu1}.
Similarly, the BELLE measurement also has the baryonium
interpretation \cite{rosner}.

\par
In the baryonium configuration of Y(4260), the constituent baryons
$\Lambda_c$ and $\bar{\Lambda}_c$ in principle should not be
restricted to being a color singlet. These two three-quark
baryon-like clusters can carry color indices, which enables the
binding energy to be easily as large as several hundred MeV, like
in the case of charmonium or bottomonium. Note that, in principle,
in the large-$N_c$ limit the baryonium binding energy can be very
large. In our case, roughly speaking, the binding energy is about
\cite{pdg}
\begin{eqnarray}
M(Y(4260)) - 2 M(\Lambda) \approx 310\; {MeV}\;.
\end{eqnarray}
Due to the large uncertainties in the experimental measurement,
these number should only be considered as an order-of-magnitude
estimation.

To distinguish our interpretation from other exotic state
explanations is easy experimentally. One of the main results of
our model is that the new observed resonance should have a large
branching ratio in the $e^+e^- \rightarrow \gamma_{ISR}
\pi^+\pi^-\psi'$ process which is suppressed in most other
schemes. Another unique feature of our model is that the two-body
decay is generally suppressed, while the three-body decay is
favored since the six-quark components of the baryonium makes the
later much easier. Recently, the BaBar Collaboration has also
measured the $e^+ e^- \rightarrow {\rm ISR} + \; p + \bar{p}$
process \cite{Babar1}. They find no clear signal of a
resonance-like structure at the energy of 4260 MeV. They set an
upper limit on the ratio of branching fractions:
\begin{eqnarray}
\frac{B(e^+ e^- \rightarrow p + \bar{p})}{B(e^+ e^- \rightarrow
\pi^+\pi^-J/\psi)} < 13\%\; .
\end{eqnarray}
In our model, this result is easy to understand. Without
considering the complicated color rearrangement, to annihilate the
charm quark pair and meanwhile to produce a pair of light quarks
there will be a hard propagator of $1/(2 m_c)^4$ in the context of
pQCD. This may suppress the $p\; \bar{p}$ production process by
about two orders of magnitude relative to the $\pi^+\pi^-J/\psi$
production.

\par
According to our point of view, the baryonium state includes no
strange quark(s) or constituent gluon(s), so its decay to the
$J/\psi$ $K^+$ $K^-$ is not favored. It would be suppressed by the
annihilation-production transition rate, which in practice may not
be too big, relative to the observed $J/\psi$ $\pi^+$ $\pi^-$
channel. However, in Refs.\cite{kou} and \cite{maiani}, by using
some subtle schemes this point can also be achieved. On the other
hand, according to Ref.\cite{li}, it is almost impossible for the
$Y(4260)$ to decay to $\psi'$ $\pi^+$ $\pi^-$ since the
$\chi_{c1}$ mass is lower than that of the $\psi'$ state. In our
model, it is possible and we can even give a rough estimation of
the relative rates of these two decay channels, i.e.,
\begin{eqnarray}
R_{\psi'/\psi} = \frac{\Gamma(Y(4260)\rightarrow
\psi'\pi^{+}\pi^{-})}{\Gamma(Y(4260)\rightarrow
J/\psi\pi^{+}\pi^{-})} \approx
\frac{\Omega_1}{\Omega_2}\frac{|\psi'(0)|^2}{|\psi(0)|^2}\approx\
0.19 \frac{\Gamma(\psi' \rightarrow e^+ e^-)}{\Gamma(J/\psi
\rightarrow e^+ e^-)}\approx 0.08\; . \label{psi}
\end{eqnarray}
Here, we neglect the dynamical difference of the new state
decaying into $J/\psi \pi^{+}\pi^{-}$ and $\psi'\pi^{+}\pi^{-}$ ,
but only integrate over kinematic phase space $\Omega_{1,2}$; the
$\psi'(0)$ and $\psi(0)$ represent the $\psi'$ and $J/\psi$ radial
wavefunctions at the origin in the quark model, respectively.
Since there are more than 100 observed $Y(4260)\rightarrow \pi^+
\pi^- J/\psi$ events, from (\ref{psi}) we would expect only a
small number of $\pi^+ \pi^- \psi'$ events with the present
experimental statistics. In addition, in our model the $e^+e^-
\rightarrow \gamma_{ISR} \pi^0\pi^0\psi$ is allowed, and the
isospin and statistical analysis tells us that it is about half
the rate of the observed one\footnote{It has also been confirmed
by the recent CLEO measurement \cite{cleo}.}, that is
\begin{eqnarray}
\Gamma[Y(4260)\rightarrow  \pi^0\pi^0\psi] \approx
\frac{1}{2}\Gamma[Y(4260)\rightarrow  \pi^+\pi^-\psi]\; .
\end{eqnarray}
While the $\pi^0\pi^0\psi$ final state is hard to pin down
experimentally, in our assignment, in principle we know this mode
should be there, contrary to the prediction in Ref. \cite{li}.

\par
In conclusion, in this work we propose a new model to explain the
recent observation of the Y(4260) state at PEP-II. In our scenario
this new resonance is a bound state of baryons, a $\Lambda_c$
pair, i.e., an ortho-baryonium state. If our assignment for the
resonance Y(4260) is correct, the pseudoscalar para-baryonium
partner should also exist in nature.

In our model, the observed nature of Y(4260) can be easily
understood. We have also made some predictions based on our
explanation for this state. Some of these predictions are
distinctively different from predictions based upon other
speculations, which are left for future experiments to measure. In
the baryonium interpretation the smallness of $K^+ K^- J/\psi$ and
non-observation of $D \bar{D}$ decay modes in the present
experiment, which are the dominant decay products in other
measured normal s-wave charmonium excited states, can be well
explained. According to our calculation, the $\psi' \pi^+ \pi^-$
decay mode of the concerned state is marginally observable with
the present experimental statistics. In our model, the decay mode
of $D^* \bar{D}^* \pi$ is observable with the collected data or in
the near future, but the mode of $D \bar{D}\pi$ is highly
restricted. And, we can explain BaBar's new upper limit for
Y(4260) to $p\; \bar{p}$. Finally, it should be noted that our
assignment for the new measured resonance (enhancement) at 4260
MeV as a baryonium state might only be an approximate description.
In reality, it could be a mixture of baryonium and convention
charmonium states.
%%%%%%%%%%%%%%%%%%%%%%%%%%%%%%%%%%%%%%%%%%%%%%%%%%%%%%%%%%%%%%%%%
\vspace{1.3cm}
\par
\par
{\bf Acknowledgments}
\vspace{.2cm}

The author is grateful for the hospitality of the theory group of
Academia Sinica, TaiWan during his visit, while the paper was
finalized. This work was supported in part by the National Natural
Science Foundation of China.
%%%%%%%%%%%%%%%%%%%%%%%%%%%%%%%%%%%%%%%%%%%%%%%%%%%%%%%%%%%%%%%%%%

\end{document}